# Disorder- and magnetic field-tuned fermionic superconductor-insulator transition in MoN thin films. Transport and STM studies.

M. Kuzmiak,[1,2] M. Kopčík,[1,2] F. Košuth,[1,3] V. Vaňo[1,2], J. Haniš[1], T. Samuely[3], V. Latyshev[3], O. Onufriienko[1], V. Komanický[3], J. Kačmarčík[1], M. Žemlička[4], M. Gmitra[1], P. Szabó[1], and P. Samuely[1]

[1]*Centre of Low Temperature Physics, Institute of Experimental Physics, Slovak Academy of Sciences, SK-04001 Košice, Slovakia*
[2]*Department of Physics, Faculty of Electrical Engineering and Informatics, Technical University of Košice, SK-04001 Košice, Slovakia*
[3]*Centre of Low Temperature Physics, Faculty of Science, P. J. Šafárik University, SK-04001 Košice, Slovakia*
[4]*QphoX B.V., Elektronicaweg 10, 2628XG Delft, Netherlands*

**Abstract**

Superconductor-insulator transition (SIT) driven by disorder and transverse magnetic field has been investigated in ultrathin MoN films by means of transport measurements and scanning tunneling microscopy and spectroscopy. Upon decreasing thickness, the homogeneously disordered films show increasing sheet resistance $R_s$, shift of the superconducting transition $T_c$ to lower temperatures with the 3 nm MoN being the last superconducting film and thinner films already insulating. Fermionic scenario of SIT is evidenced by applicability of the Finkel'stein's model, by the fact that $T_c$ and the superconducting gap $\Delta$ are coupled with a constant ratio, and by the spatial homogeneity of the superconducting and electronic characteristics. The logarithmic anomaly found in the tunneling spectra of the non-superconducting films is further enhanced in increased magnetic field due to the Zeeman spin effects driving the system deeper into the insulating state and pointing also to fermionic SIT.

*Magnetotransport, superconductor-insulator transition, STM, tunneling spectroscopy, strongly disordered ultrathin superconducting MoN films*

## 1 Introduction

A quantum phase transition from superconducting to insulating state (SIT) can be induced in quasi two-dimensional systems by disorder or magnetic field. This transition occurs at the critical disorder, where the electron's mean free path is extremely small and the Ioffe-Regel parameter, the product of the Fermi momentum and the mean free path $k_Fl$ is close to unity [1,2]. We distinguish two SIT mechanisms depending on whether the growing disorder affects the amplitude or the phase of the superconducting wave function [2]. At the loss of long-range phase coherence superconducting condensate melts into locally coherent superconducting islands which can survive even in the insulating state [3]. This is termed a bosonic SIT. A fermionic SIT occurs when disorder reduces the electronic screening of the Coulomb interaction and the amplitude of the wave function is suppressed. Then, transition temperature $T_c$ and superconducting order parameter/gap $\Delta$ disappear at once and a bad metallic phase is formed before a transition to the insulating state at even higher disorder [4].

The bosonic SIT has been observed by scanning tunneling spectroscopy in TiN, InO$_x$, NbN and $a$-MoGe showing emergence of local superconducting puddles with varying superconducting gap present on both superconducting and insulating sides of the transition and a pseudogap above critical temperature $T_c$ or magnetic field $B_{c2}$ [5–11]. Loss of phase coherence can also be manifested by variation of the coherence peaks height of the superconducting tunneling spectra among individual puddles [9,12,13]. Recent scanning tunneling microscopy (STM) experiments of Dutta et al. [14] performed on ultrathin $a$-MoGe films near critical disorder show a Bose (bad) metallic state above the critical field still revealing traces of superconducting gap. Then, in both scenarios quantum transition from superconducting to bad metallic state may happen before the insulating state is achieved.

At fermionic SIT when increased Coulomb interaction is operative, neither pseudo-gap above $T_c$ or $B_{c2}$, nor spatial variation of the superconducting gap appear, i.e. "fermionic" systems are spatially homogeneous at homogeneous disorder. Recently, Carbillet et al. [15,16] showed that inhomogeneities of the gap on the scale much larger than the superconducting coherence length $\xi$ can appear in the fermionic SIT as well. This is due to emerging inhomogeneities of the normal state where the strength of the Altshuler-Aronov effect varies. Early tunneling experiments performed on planar tunnel junctions suggested a fermionic mechanism for the SIT in Bi and PbBi/Ge thin films [17,18]. In our works on disordered ultrathin MoC films [19,20] we presented a direct evidence that $T_c$ and $\Delta$ are simultaneously suppressed and superconductivity remains spatially homogeneous very close to SIT. When SIT was driven by magnetic field oriented perpendicularly to the heavily disordered but still superconducting MoC film, surprisingly, the Zeeman paramagnetic effects dominated over orbital coupling on both sides of the transition [20].

Here, we present studies on MoN films starting from 30 nm thickness down through the whole SIT where 3 nm MoN is the last superconducting film and thinner films no longer exhibit superconductivity on large scale as evidenced by resistive measurements. Based on our transport and low-temperature STM measurements we show that the SIT in MoN thin films tuned by both disorder and/or magnetic fields is realized by the fermionic pathway. In particular, the spin-dependent AA effect in insulating state points to the dominance of Zeeman effects supporting the fermionic scenario.

## 2 Experimental

Molybdenum forms several nitrides including cubic γ-phase of $Mo_2N$, tetragonal β-$Mo_2N$ and hexagonal δ-MoN [21]. The γ-phase has a superconducting transition at about 5 K [22,23]. Recently, Tsuneoka *et al.* [24] prepared series of thin γ-phase $Mo_2N$ films down to 1.5 nm thickness showing the SIT. Here, we have prepared the high quality $Mo_2N$ thin films with $t$ = 30, 10, 5, 3, 2, 1.5 nm thicknesses by reactive magnetron sputtering onto a sapphire substrate from a Mo target in an argon-nitrogen mixture gas atmosphere. The crystal structure of the films was characterized by X-ray diffraction. Our films reveal a face-centered cubic crystalline structure, which is typical for a γ-$Mo_2N$ phase [25,26].

The basic characterization of our samples was done by transport measurements. A more detailed discussion can be found in the Suppl. Mat. [27], where the determination of the Ioffe-Regel parameter [1] is explained and the analysis of our data in terms of the Finkel'stein's model [4] is also described. We applied scanning tunneling microscopy (STM) measurements to study the quasiparticle density of states (DOS) of our samples. The experiments were performed in a homemade STM system immersed in a Janis SSV $^3$He cryostat, which allows measurements down to $T$ = 0.3 K in magnetic fields up to $B$ = 8 T. A gold STM tip was used to measure the ex-situ prepared samples. At each point of the sample surface, first a topography was acquired with the feedback loop on, then the feedback loop was turned off and the tunneling current was recorded while sweeping the bias voltage, all on the grid of 128x128 topography points.

The tunneling conductance was calculated by numerical differentiation of the *I-V* curves. The *dI(V)/dV* spectra measured through the tunnel junction between a normal metal and a superconductor can be described by the equation

$$\frac{dI(V)}{dV} = N \int_{-\infty}^{\infty} N_S(E) \left[ \frac{-\partial f(E+eV)}{\partial eV} \right] dE, \quad (1)$$

where $E = eV$ is the energy and the derivative of the Fermi distribution in the square brackets represents thermal smearing. $N$ is the constant DOS of the metallic tip and $N_S$ represents the superconducting DOS, resp. The BCS theory defines $N_S$ as

$$N_S(E) = \mathrm{Re}\left(\frac{E}{\sqrt{E^2 - \Delta^2}}\right), \quad (2)$$

where $\Delta$ is the superconducting energy gap. Tunneling spectra with spectral smearing higher than the thermal broadening can be described by the phenomenological modification of the BCS DOS of Dynes [28] who introduced a complex energy $E = E'-i\Gamma$, where $\Gamma$ is the spectral smearing parameter. The non-zero value of this parameter can be connected with pair-breaking centers at the tunnel junction area [29]. If the superconductor is in the normal state and $\Delta = 0$, the DOS will be constant in the case of standard metals. However, in the case of highly disordered metals, we must account for corrections related to the enhanced electron-electron interaction. According to the Altshuler-Aronov (AA) theory of 2D systems [30,31] this is manifested by a logarithmic reduction of the normal state DOS at Fermi energy down to the thermal energy defined by the measuring temperature. The standard AA correction is not a function of the applied magnetic field. However, when the electron spins are taken into account in the particle-hole channel the AA correction to the normalized DOS $N_N(E) = 1 - \delta N$ reveals field dependence, described as [31]

$$\delta N(E, T, B) \approx A \left\{ f(E, \Gamma_0) \right.$$
$$+ \frac{\lambda_1}{2\lambda_0} \left[ f(E, \Gamma_0) \right. \quad (3)$$
$$\left. \left. + \frac{1}{2} \sum_{\alpha=\pm 1} f(E + \alpha E_Z, \Gamma_0) \right] \right\}$$

where *A* includes all constants and parameters of the tunneling barrier, *f* is the logarithmic function defined in [20] with arguments normalized to the thermal energy and $E_Z = 2\mu_B B$ is the Zeeman energy, with $\mu_B$ being the Bohr magneton and *B* magnetic field. The constants of $\lambda_0$ and $\lambda_1$ describe the interaction of an electron and a hole with total spin 0 and 1, respectively and $\Gamma_0$ is responsible for the broadening of the AA logarithmic singularity [20]. The first and the second terms originate from the interaction of the electron and the hole with a zero projection of the total spin on the direction of the magnetic field, and the third (and fourth) term from interactions with opposite spins. In the case when the ratio $\lambda_1/2\lambda_0$ is from interval $\langle -1/3, 0 \rangle$ at zero magnetic field the logarithmic anomaly of the first term is partly compensated by three equal logarithmic terms with opposite sign. In the presence of applied magnetic field a part of the electronic DOS at the Fermi energy is removed due to spin polarization and two singularities/maxima should appear at $E = \pm E_Z$. Then, the logarithmic singularity/minimum at the Fermi level is more pronounced than it was in the zero-field case.

## 3 Results

The temperature dependencies of the sheet resistance $R_s(T)$ obtained on 30, 10, 5, 3, 2 and 1.5 nm thin MoN films at temperatures from 100 K down to 0.3 K are shown in Fig. 1a). The thickest 30 nm sample shows an almost temperature independent sheet resistance down to the superconducting transition at $T_c = 5.83$ K. With reducing film thickness the normal state sheet resistance $R_s$ increases, the derivative $dR_s/dT$ becomes negative and a gradual shift of the superconducting transition typical for homogenously disordered systems is visible down to $T_c = 2.8$ K in the case of the 3 nm sample. The evolution of this disorder-tuned superconductor-insulator transition can be followed in Fig. 1b), where the critical temperature $T_c$ (blue symbols - left scale) and the Ioffe-Regel parameter $k_Fl$ (green symbols - right scale) are shown as the function of the film thickness. The inset of Fig. 1b) shows $T_c$ as a function of the sheet resistance $R_s$ by open black symbols. The solid red line is the fit to the Finkel'stein's formula [4] with the fitting parameters of the critical resistance $R_{cr} = 2.3$ kΩ where the superconductivity terminates, the critical temperature of the bulk phase $T_{c0} = 6.34$ K and the elastic scattering time $\tau = 2.66 \times 10^{-16}$ s. The rapid decrease of the $k_Fl$ values with reducing film thickness and the low value of $k_Fl \sim 1.4$ determined at the 3 nm sample suggest that this superconducting sample is in the near proximity to critical disorder. From transport measurements in transverse magnetic field the upper critical magnetic field was also determined, see Fig. S1 and S2 in Suppl. Mat. [27].

Low temperature STM measurements were performed to study the surface topography of our thin films. Figure S3 presented in the Suppl. Mat. [27] shows typical topography images of the MoN thin films from both sides of the SIT. The samples had clean surfaces with continuous and compact polycrystalline structure. The images shown in Fig. S3 [27] can be characterized by the average surface corrugations ~ 1.2 nm, 0.5 nm and 0.35 nm for the 30 nm, 3 nm and 2 nm sample, respectively. The sharp, well defined grain boundaries and the low averaged surface corrugations point to high quality of the studied thin films.

In Fig. 2, representative temperature dependencies of the STM tunneling spectra measured on 30, 3, and 2 nm samples are shown in panels a), b), and c), respectively. The upper panels show 3D plots of the zero-field temperature dependencies of the $dI(V)/dV$ spectra. The spectra taken on 30 nm and 3 nm samples reveal standard superconducting features. At increased temperature the superconducting gap structure bordered by symmetrical coherence peaks is suppressed at $T_c$ and at higher temperatures constant normal state tunneling spectra are visible. The value of the superconducting energy gap $\Delta$ and its temperature dependence in both sets of the samples have been determined from fitting to Dynes modification of the BCS DOS. The fitting curves (open symbols) and experimental data (lines) at the indicated temperatures are shown in the corresponding lower panels. We can see clearly that the model describes the experimental data perfectly. The obtained temperature dependencies of the energy gap are shown in Fig. S4 of the Suppl. Mat. [27]. The gap values reveal BCS-like temperature dependencies with $\Delta(0) = 0.87$ meV, $\Gamma/\Delta \sim 0.06$ meV, $T_c = 5.6$ K, $2\Delta(0)/k_BT_c \sim 3.6$ for 30 nm and $\Delta(0) = 0.44$ meV, $\Gamma/\Delta \sim 0.4$ meV, $T_c = 2.9$ K, $2\Delta(0)/k_BT_c \sim 3.55$ for the 3 nm sample. Figure 2c) shows the effect of temperature on the tunneling spectrum measured on the non-superconducting 2 nm film. No coherence peaks indicative of superconductivity are observed at the lowest temperature $T = 0.5$ K but the spectrum reveals a logarithmic-like reduction towards zero-bias voltage. We fit our experimental data to the thermally smeared AA correction, defined in (3). The curve taken at the lowest temperature is fit by parameters: $A = -0.26$, $\lambda_1/2\lambda_0 = -0.12$, $\Gamma_0 = 0.15$ meV and $B = 0$ T. In the fits at higher temperatures, only the temperature was changed and the smearing parameter varied slightly in the range of 5 %. The agreement between our data (lines) and the thermally smeared AA correction (symbols) fit indicates, that what we observe is not the temperature dependence of the logarithmic reduction, but only the effect of increased thermal smearing with energy $\sim k_BT$.

We also studied the effect of perpendicular magnetic field on the tunneling spectra on samples with thicknesses of 30 nm, 3 nm, 2 nm and 1.5 nm at $T = 0.5$ K. The magnetic dependence of the $dI(V)/dV$ spectra measured on the thickest 30 nm superconducting sample shown in Fig. S5 in Suppl. Mat. [27] reveals a classical behavior. In increasing field, the position of gap-like peaks shifts to lower voltage and the gap gradually closes. This process can be better followed in the side view shown in the inset of Fig. S5 and in top view of Fig. S6a) in Suppl. Mat. [27]. The upper critical magnetic field of the 30 nm sample is above the range of our experimental set-up (Fig. S2 in Suppl. Mat. [27]), so we could not detect a full transition to the normal state.

The main panels at the top of Fig. 3 show 3D plots of the effect of perpendicular magnetic field on the tunneling spectra on the 3 nm, 2 nm and 1.5 nm thin films. The insets show side views of the corresponding magnetic dependencies. Fig. 3a) plots the effect of magnetic field on the thinnest (3 nm) superconducting sample that is close to the critical disorder. Again, at increasing field the gap peaks are getting more smeared and the in-gap area is gradually filled but remarkably, the gap-peak position remains unchanged and before the conductance $dI(V)/dV$ gets completely flat at the upper critical magnetic field $B_{c2} \sim 5$ T, the "gap" structure holds and is even enhanced in further increased field. The both effects are also documented in Fig. S6 b) of Suppl. Mat. [27]). We have seen similar effects of the magnetic field on ultrathin MoC films in our previous paper [20], where we studied the magnetic field induced SIT in these samples. The both unusual features were described by the Zeeman effect on the density of states, namely, the constant position of the gap peak as due to spectral broadening of the Zeeman split superconducting gap peaks and enhanced gap in the normal state due to Zeeman effect on spin dependent AA correction to DOS. There, based on the work of [29], to describe the total DOS we used the product of $N(E) = N_S^D(E)N_N(\Omega)$, where $N_S^D$ is the Dynes modification of the BCS DOS (2) with smearing parameter $\Gamma$, $N_N(\Omega)$ represents spin dependent AA corrections defined in (3), and the energy-dependent parameter $\Omega(E) = \mathrm{Re}\left[\sqrt{(E+i\Gamma)^2 - \Delta^2}\right]$ ensures the balance of quasiparticles after the transition to the superconducting state. Here, we employ this model to describe the magnetic field dependence of the tunneling spectra measured on our 3 nm thin MoN superconducting sample. We started to fit our spectra in the normal state at $B = 8$ T without superconducting contribution. We determined the values of the parameters $A = -0.3$, $\lambda_1/2\lambda_0 = -0.24$, $\Gamma_0 = 0.31$ meV which we used to fit the complete magnetic dependence down to $B = 0$ T. During the fit in the superconducting state at $B < B_{c2}$ we used only the value of the energy gap as a fit parameter, which we varied according to formula $\Delta = \Delta(0)\sqrt{1 - B/B_{c2}}$, in agreement with our results on MoC [20]. The bottom panel of Fig. 3a) shows the results of our fit (open symbols) together with selected experimental curves from the upper panel (solid lines). The magnetic field dependence of the AA correction used during the fit is shown in Fig. S7 of the Suppl. Mat. [27]. It can be seen that the small AA correction at $B = 0$ T with the ZBC value of $\sim 0.95$ is enhanced by $\sim 20$ % to the value of $\sim 0.75$ at $B = 8$ T.

While in the 3 nm MoN film and similarly in MoC films in our previous study [20] superconductivity and AA effect interfere, in the thinner MoN samples one can focus on the AA effects only. As can be seen in Fig. 3b) and c) the AA effect (reduction of the DOS at Fermi energy) gets stronger on the thinner and more disordered 1.5 nm sample compared to 2 nm MoN film. The most importantly, also the amplifying effect of magnetic field on the AA reduction is unambiguously detected, here: the zero-bias conductance drops by roughly $\sim 10$ % from $\sim 0.67$ on 2 nm film and $\sim 4$ % from 0.52 for the 1.5 nm film when the magnetic field is increased from $B = 0$ to 8 T.

To describe these magnetic field dependencies quantitatively, we fit the spectra at selected magnetic fields as shown in the lower panels of Fig. 3b) and c). Since these samples are not superconducting, the superconducting contribution from the total density of states falls out, and we can fit the tunnel spectra to the DOS defined as $N_N(E) = 1 - \delta N$ according to formula (3). We started to fit the spectra again at the highest field $B = 8$ T, where we obtained the following fitting parameters: $A = -0.26$, $\lambda_1/2\lambda_0 = -0.12$, $\Gamma_0 = 0.17$ meV for the 2 nm sample and $A = -0.23$, $\lambda_1/2\lambda_0 = -0.035$, $\Gamma_0 = 0.14$ meV for the 1.5 nm sample. At lower magnetic fields only the value of $\Gamma_0$ was slightly corrected by less then 3 %.

As a result, the spin dependent AA correction to the normal density of states describes well the effect of magnetic field in the both non-superconducting samples, and also in our thinnest 3 nm superconducting sample. In the latter case remarkably, the field dependent AA correction contributes very weakly ($< 4$ %) to the total DOS in the superconducting state at $B < B_{c2}$ (Fig. S7 in Suppl. Mat. [27]).

In Fig. 4, we compare the AA effect in zero magnetic field in 3 nm, 2 nm, and 1.5 nm samples. In the case of 3 nm sample, we plot only the fitting curve of the contribution of the AA effect because in the experiment the total DOS comprises also the superconducting part. One can see a rapid enhancement of the AA correction upon decreasing the thickness/increasing disorder which is even amplified by the fact that spin dependent contributions partly compensating the leading AA term are also diminished with disorder.

## 4 Discussion

In the following we argue that the SIT introduced in our thin MoN films by disorder and magnetic field follows the fermionic scenario.

*First* argument is based on the fact that superconducting transition temperature $T_c$ versus the films sheet resistance $R_s$ in Fig. 1 follows the predictions of the Finkel'stein's model based on enhanced Coulomb effects leading to suppression of the amplitude of the order parameter and $T_c$ at once [4]. This finding has also been documented in our transport measurements on strongly disordered MoC films [19] and in other materials as for example MoGe films [32]. Tsuneoka et al. [24] who prepared a series of MoN films from 60 nm down to 1.5 nm with $T_c$ decreasing from 6.6 K down to 2.5 K and the SIT transition at about 5 nm thin films also argued that $T_c$ depression obeys the fermionic scenario according to the amplitude fluctuation in the order parameter. They could fit $T_c$ as a function of the sheet resistance to the Finkel'stein's model with the following parameters: the bulk $T_{c0} = 6.45$ K, the critical resistance $R_{cr} = 2$ k$\Omega$ and the elastic scattering time $\tau = 1.6 \times 10^{-16}$ s values very close to ours. It should be noted that the suppression of $T_c$ with increasing value of $R_s$ has also been observed in systems with an evident bosonic SIT, hence the validity of the Finkelstein model is a necessary but not a sufficient condition of the fermionic model [11].

*Second* evidence supporting the fermionic SIT is based on the fact that with increasing disorder the superconducting energy gap $\Delta$ and the transition temperature are suppressed simultaneously and the superconducting coupling is constant, not depending on the increasing disorder. In our case it is so and the coupling parameter is $2\Delta(0)/k_BT_c \sim 3.5 - 3.6$ for the 30 nm film down to 3 nm MoN (see Fig. S4 in [27]). This has also been observed in our previous paper in MoC films [19].

*Third* argument employs the fact that spectral characteristics on increasingly disordered superconductors remains homogeneous, i.e. a long-range coherence holds and no local variations of the superconducting gap emerge. We have studied the superconducting energy gap locally by the STM measurements on all our superconducting samples down to 3 nm thickness close to the critical disorder. The gap maps constructed from these measurements, as well as the zero-bias conductance maps (ZBC) demonstrate the homogeneity of the superconducting state on the flat regions of the investigated surfaces. Homogeneity of tunneling spectra measured on the 3 nm sample along a 40 nm white line from Fig. 5a) is presented in the form of top view on spectra in Fig. 6a).

We also studied the coherence-peaks height map on the 3 nm sample, which could reflect the loss of phase coherence, too [9,12,13]. Before constructing this map, we normalized all tunneling spectra to the contribution of the AA effect at $B = 0$ T (black symbols in Fig. 4) determined from fit in Fig. 3. In contrast to experiment on NbN [9] we could not find any distinct variations in the coherence peak heights. On the other hand, we admit that with the high spectral smearing with $\Gamma/\Delta \sim 0.4$ it would be difficult.

In some areas of the 3 nm thin superconducting film, islands a few nanometers wide with slightly different values of the superconducting energy gap than in other locations are observed. However, as already shown in our previous paper [33] such a variation occurs at the places with slightly more pronounced protrusions/thicknesses of the film and a good correlation exists between the topography and the gap maps. The thicker the area the bigger the energy gap $\Delta$ and the smaller the spectral smearing $\Gamma$. As we have already shown on MoC disordered films [34] the effect is connected with the pair-breaking fields at the interface between the substrate and the thin film.

Obviously, the question may arise how close to SIT the disordered films would have to be for bosonic effects to occur, and whether our 3 nm MoN was that close. For example, in strongly disordered NbN [10] and $\alpha$-MoGe [11] thin films, samples thicker than 8 nm and 5 nm, resp., exhibit an evident fermionic scenario with a spatially homogeneous BCS-like gap, but on thinner samples with higher disorder, spatial inhomogeneities appear. Therefore, we study here for the first time not only strongly disordered superconducting samples near SIT but also non-superconducting MoN films that are already on the other side of the transition. Remarkably, the spectral maps taken on the 3 nm samples above $B_{c2}$ and the spectral maps taken on the non-superconducting 2 nm and 1.5 nm samples measured at any field show no spatial variations regardless of any corrugations. While for the 3 nm MoN film in superconducting state the zero-bias conductance map taken at zero field in Fig 5b) reveals small variations in agreement with the topography in Fig 5a) (see for example the spots in the white ovals, where an increase of about 1 nm in the topography corresponds to a 30 - 50 % lower ZBC due to locally bigger superconducting gap), on the contrary, the ZBC map in c) taken at 7.5 T, i.e. above $B_{c2}$, features no areas correlated with topography, only noise with approx. 5 % dispersion. In Fig. 5d), the 2 nm MoN topography shows higher and lower spots, but no spatial changes are observed in the ZBC maps measured at 0 T and 5 T (see e) and f), respectively) except experimental noise. Similar ZBC maps without apparent spatial changes were also observed on the 1.5 nm sample. The 40 nm long linecuts, measured on the surfaces of the superconducting 3 nm thin film in the normal state above $B_{c2}$ and non-superconducting 2 nm and 1.5 nm thin films at $B = 0$ T shown in Fig. 6 b), c), d), respectively, also point to the homogeneity of the measured spectra.

Carbillet et al. [15,16] pointed out that inhomogeneities of the superconducting gap can appear due to local variations of the Altshuler-Aronov effect. They modeled the AA correction as $dI/dV \sim V^{\alpha}$ and showed the local variations in the $\alpha$-maps on a rather long scale. We have also analyzed the spectral maps measured on 3 nm (at 7.5 T), 2 nm and 1.5 nm MoN films at different fields on surfaces typically $50 \times 50$ nm$^2$ in a similar way but our $\alpha$-maps did not show any observable variations. However, we admit that it is not ruling out their existence completely, because of the small size of the studied surfaces and high spectral smearing of the MoN spectra.

*Fourthly*, we show, that the transverse magnetic field applied to the thin films in close proximity to SIT acts through paramagnetic Zeeman spin-splitting effect. In the previous work [20] on 3 nm superconducting MoC film paramagnetic Zeeman effects dominated on both sides of the field-tuned SIT. The critical magnetic field $B_{c2}(T)$ showed square root temperature dependence typical for paramagnetic pair breaking, the Zeeman effect was found in the superconducting density of states below $B_{c2}$, and at higher fields, in the normal state a depleted DOS due to Altshuler-Aronov effect was detected. Also, the logarithmic AA anomaly in DOS present in the STM spectra at $B_{c2}$ was enhanced upon increasing magnetic field driving the system deeper into the insulating state. The MoN films close to SIT indicate very similar effects. The spin-dependent and spatially homogeneous AA effect on superconducting 3 nm and thinner non-superconducting samples is identified. Obviously this can be observed only in systems where the magnetic field acts on quasiparticles through Zeeman splitting. Therefore, it is very important that the magnetic field dependence of the superconducting energy gap $\Delta(B) \sim \sqrt{1 - B/B_{c2}}$ typical for the presence of paramagnetic pair breaking of Cooper pairs was necessary in the fits on the 3 nm superconducting sample. The temperature dependence of the critical magnetic field $B_{c2}(T)$ of our sample shown in Fig. S2 [27] also indicates the dominance of paramagnetic effects. This follows not only from the square-root curvature of $B_{c2}(T)$ in the vicinity of $T_c$ (red dashed line) but also from the high value of $B_{c2}(0) \sim 4.5 - 5$ T close to the Clogston limit ($B^P[\text{T}] \simeq 1.8 T_c[\text{K}]$) [35], which for the superconductor with $T_c = 2.9$ K is $B^P = 5.2$ T. The determination of the $B_{c2}$ values from our transport measurements is described in detail in the Suppl. Mater. [27]. Thus, in strongly disordered MoN thin films, when a transverse magnetic field is applied, Zeeman spin effects dominate both in the superconducting and the normal states.

The dominance of spin splitting effects has also been observed in systems that are close to the metal-insulator transition and exhibit "bad metallic" behavior due to strong electron correlations. Bogdanovich *et al.* [36] have shown that in boron-doped Si:B samples, the metal-insulator transition is controlled predominantly by electron correlations and orbital effects are unimportant despite strong spin-orbit effects.

## 5 Conclusions

MoN films reveal a transition from superconducting to insulating/bad metallic phase upon decreasing thickness from 30 nm to 1.5 nm. The superconducting transition temperature is gradually shifted from about 6 K to 2.7 K in the case of 3 nm films accompanied by decreasing Ioffe-Regel parameter approaching unity. Films thinner then 3 nm are already non-superconducting. The shift of $T_c$ is also accompanied by the increase of the sheet resistance and can be modelled by the Finkel'stein's model for the enhanced electron interaction providing the critical sheet resistance $R_{cr} = 2.3$ kΩ where the superconductivity terminates. The scanning tunneling microscopy and spectroscopy measurements prove that $T_c$ and the superconducting gap Δ are coupled with a constant ratio for all films and that the superconducting DOS and the normal-state Altshuler-Aronov DOS corrections are spatially homogeneous and no emergent granularity appears on either sides of SIT. The logarithmic AA DOS correction, is enhanced in increased magnetic field due to the Zeeman spin effects driving the system deeper into the insulating state. All this is pointing to purely fermionic SIT scenario in the system.

**Acknowledgments** We gratefully acknowledge helpful conversations with M. Grajcar and R. Hlubina.

**Funding Information** This work was supported by the projects APVV-20-0425, APVV-20-0324, VEGA 2/0058/20, VEGA 1/0743/19, IMPULZ IM-2021-42, the EU ERDF Grant No. VA SR ITMS2014+ 313011W856, by the European Union's Horizon 2020 Research and Innovation Programme under Grant Agreement No. 824109 (European Microkelvin Platform), the COST action CA21144 (SUPERQUMAP) and by U.S. Steel Košice.

**Figures:**

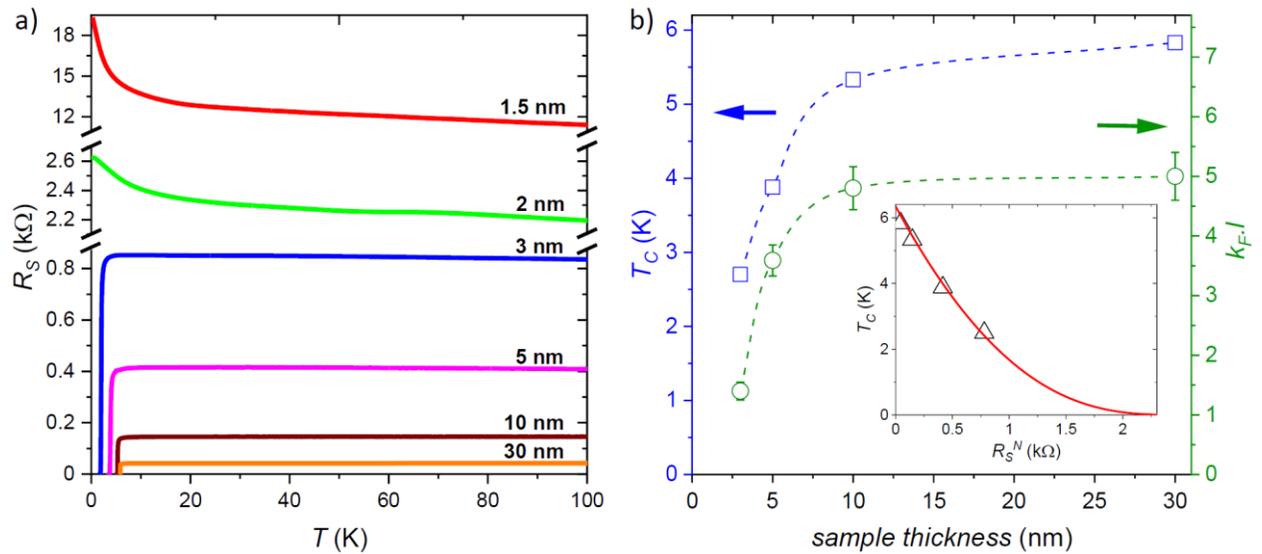

**Fig. 1:** a) Evolution of the temperature dependencies of the sheet resistance $R_S(T)$ with various thicknesses of MoN thin films. Film thicknesses are shown above the respective curves. b) The main panel shows thickness dependencies of the critical temperature $T_C$ (blue color, symbol □) and the Ioffe-Regel parameter $k_F.l$ (green color, symbol ○). The $T_C$ values have been determined at 50 % of the normal state sheet resistance. The inset shows the dependence of the normal state sheet resistance $R_S^N$ on the critical temperature $T_C$ (black △ symbols). The red curve is a fit to the Finkel'stein's model with fitting parameters given in the legend. The analysis of the transport data is detailly described in the Suppl. Mat. [27].

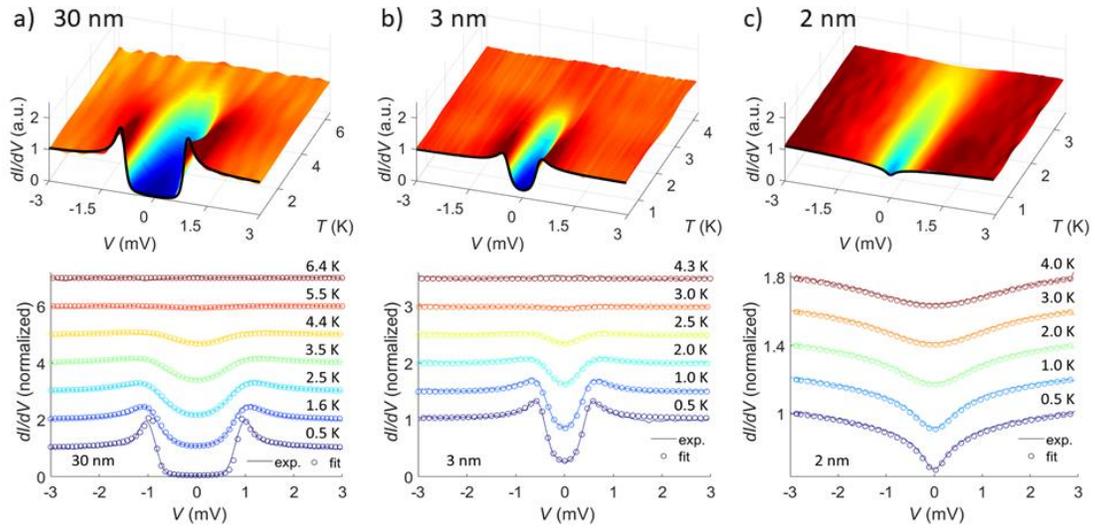

**Fig. 2**: color plots of the typical STM tunneling spectra measured on MoN thin films with 30, 3 and 2 nm thicknesses at different temperatures at $B = 0$ T a), b), c) are shown in the upper panels. All spectra have been normalized to their values at 3 mV bias. The tunneling spectra from a) and b) were fitted to Dynes formula, spectra from c) to Altshuler-Aronov effect defined in formula (1) including thermal smearing in all cases. Selected experimental curves from the upper panels (lines) and the fitting curves (symbols) are shown in the corresponding lower panels.

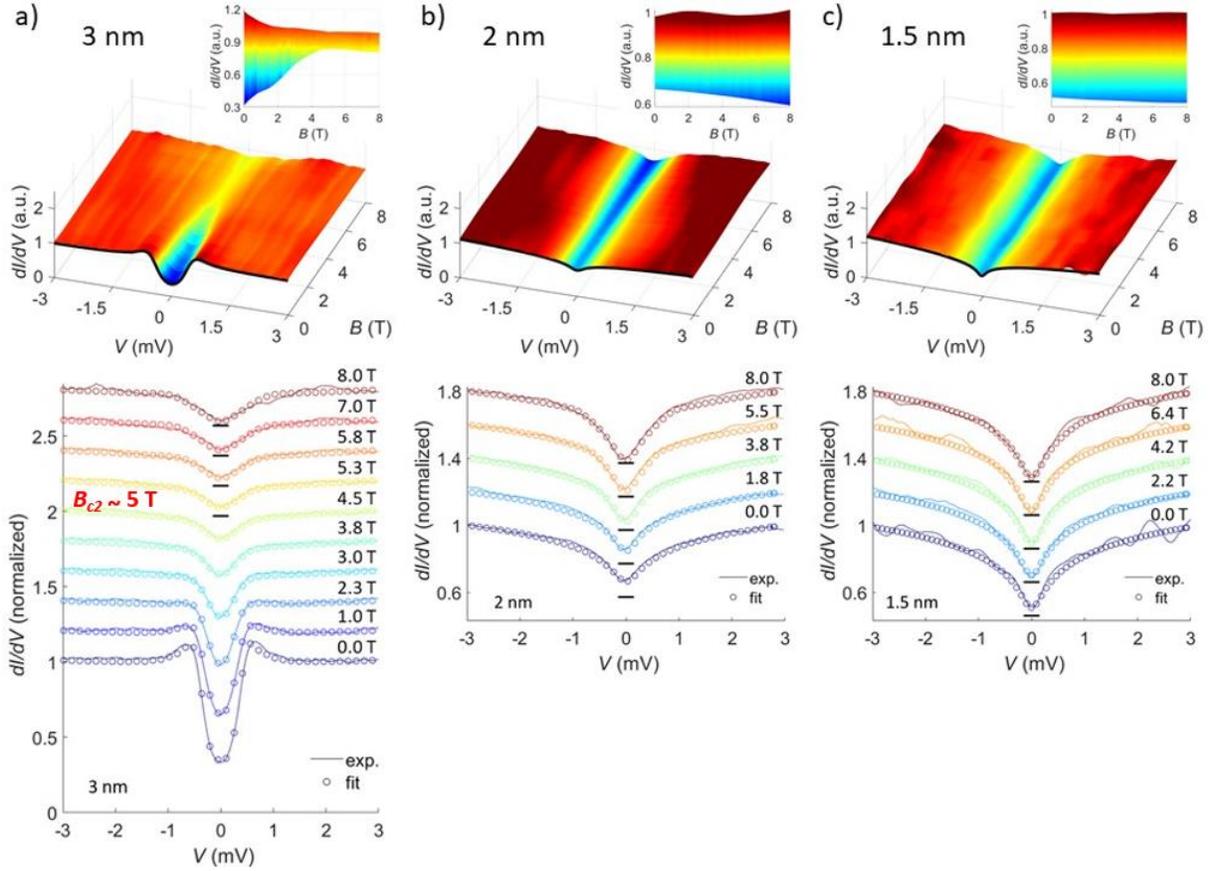

**Fig. 3**: color plots of the typical STM tunneling spectra measured on MoN thin films with 3, 2 and 1.5 nm thicknesses at different magnetic fields at $T = 0.5$ K a), b), c) are shown in the upper panels. All spectra have been normalized to their values at 3 mV bias energy. The magnetic field dependencies from a) and b) correspond to spectra shown in Fig. 2b) and c). The tunneling spectra from a) were fitted to Dynes formula including Altshuler-Aronov corrections, spectra from b) and c) to Altshuler-Aronov effect defined in formula (3) including thermal smearing in both cases. Selected experimental curves from the upper panels (lines) and the fitting curves (symbols) are shown in the corresponding lower panels. The horizontal black lines at each set are positioned to the conductance values (at $B = 8$ T) 0.75, 0.6 and 0.48 for 3, 2 and 1.5 nm thin films, respectively.

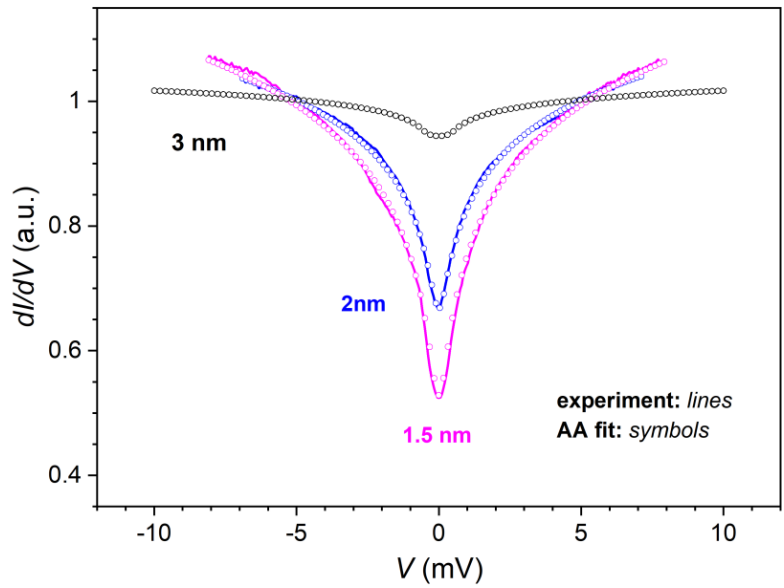

**Fig. 4:** Typical tunneling spectra measured in the non-superconducting samples with 1.5 and 2 nm thicknesses (pink and blue line, resp.) at $T = 0.5$ K. The fitting curves to Altshuler-Aronov model are shown with symbols of corresponding color. The black symbols show the theoretically calculated Altshuler-Aronov contribution to the total DOS at the superconducting 3 nm sample at $B = 0$T. All curves were normalized to their 5 mV bias values.

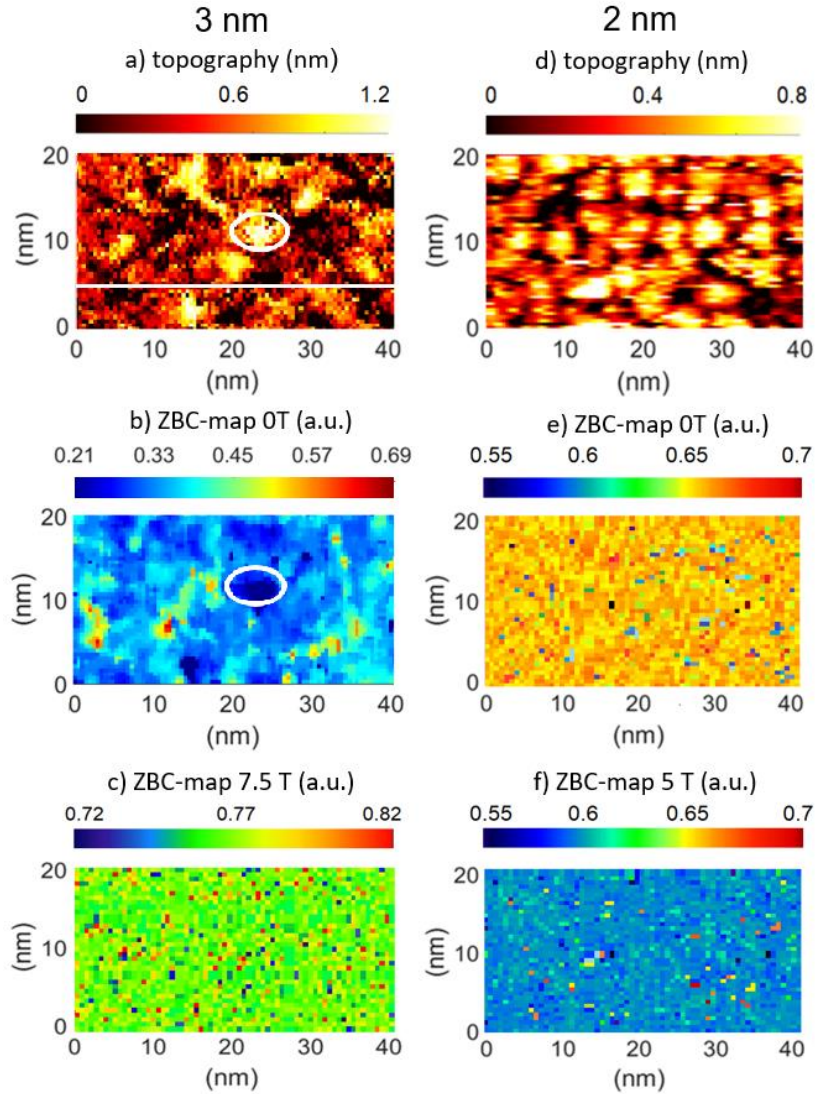

**Fig. 5:** Surface topographies a), d) and the corresponding Zero-Bias-Conductance (ZBC) maps b), c), e), f) of the superconducting 3 nm (left panels) and non-superconducting 2nm (right panels) MoN thin films taken at indicated magnetic fields. 3 nm sample shows protrusions as this in the oval in a) which is reflected also in deeper ZBC data in the oval in b) taken in superonducting state at $T = 0.5$ K and $B = 0$ T while above $B_{c2}$ this contrast is lost, see c). Protrusions found in fully non-superconducting 2 nm sample in d) has no reflections in either of ZBC maps in e) and f).

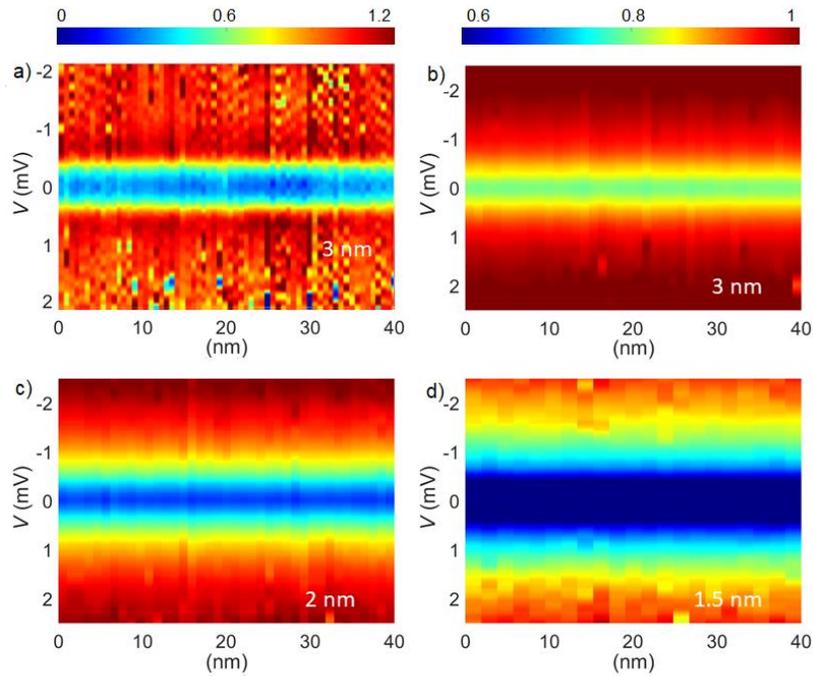

**Fig. 6:** Typical 40 nm long linecuts of the tunneling spectra constructed from the spectral maps measured on 3 nm thin film in the superconducting ($T = 0.5$ K and $B = 0$ T) and normal state ($B = 7.5$ T $> B_{c2}$) in a) and b), respectively, and on non-superconducting 2nm and 1.5nm thin films ($T = 0.5$ K and $B = 0$ T), c) and d), respectively. Tunneling spectra in a) are taken along the white line in Fig. 5a). Figures b)-d) are plotted in the same color scale, shown above b).

# Supplemental Material

# Disorder- and magnetic field-tuned fermionic superconductor-insulator transition in MoN thin films: Transport and scanning tunneling microscopy

M. Kuzmiak,[1,2] M. Kopčík,[1,2] F. Košuth,[1,3] V. Vaňo[1,2], J. Haniš[1,3], T. Samuely[3], V. Latyshev[3], O. Onufriienko[1], V. Komanický[3], J. Kačmarčík[1], M. Žemlička[4], M. Gmitra[1], P. Szabó[1], and P. Samuely[1]

[1]Centre of Low Temperature Physics, Institute of Experimental Physics, Slovak Academy of Sciences, SK-04001 Košice, Slovakia
[2]Department of Physics, Faculty of Electrical Engineering and Informatics, Technical University of Košice, SK-04001 Košice, Slovakia
[3]Institute of Physics, Faculty of Science, Pavol Jozef Šafárik University in Košice, SK- 04001 Košice, Slovakia
[4]QphoX B.V., Elektronicaweg 10, 2628XG Delft, Netherlands

**Magnetotransport measurements**

The magnetotransport measurements have been performed with a standard DC four-probe technique, where four 50 μm silver wires were attached with silver paste to the edges of the 5 × 5 mm$^2$ sample The magnetic field was applied perpendicular to the plane of the MoN thin films. The sheet resistance $R_S$ as a function of temperature $T$ was measured down to 1.4 K using an Oxford Instruments VTI system, at lower temperatures down to $T$ = 0.3 K in PPMS and Janis SSV $^3$He refrigerator. Fig. S1 shows our magnetotransport data obtained on 30 nm a), 10 nm b) and 3 nm c) thin films at magnetic fields marked in the legends. The upper critical magnetic field $B_{c2}$(T) was determined at 90% of the normal state sheet resistances of the $R_S(T)$ curves measured at fixed magnetic fields (Fig. S2). While $B_{c2}$(T) curves of the 30 nm and 10 nm thin films show conventional linear temperature dependencies near $T_c$ typical for orbital pair breaking, the critical field of the 3 nm thin film reveales a negative curvature with a square root dependence $\propto \sqrt{T_c - T}$ (red dashed line). The change in the curvature of the $B_{c2}$(T) dependencies near $T_c$ from linear to square root indicate that in our samples with increasing disorder the pair-breaking mechanism changes from orbital (30 nm and 10 nm films) to paramagnetic (3 nm film) [*Y. Matsuda and H. Shimahara, J. Phys. Soc. Jpn. 76, (2007) 051005*]. To confirm we also applied a different method, the well-known Ullah-Dosey scaling procedure for 2D superconductors to determine the $B_{c2}$ values on the 3 nm thin films [*S. Ullah and A.T. Dorsey, Phys. Rev. Lett. 44 (1991) 262*]. Ullah and Dorsey (UD) proposed a scaling of the superconducting fluctuation conductivity related to the fluctuations of the amplitude and the phase of the order parameter. Then the fluctuation conductivity $\Delta\sigma_{2D}$ scales as

$$\Delta\sigma_{2D}^{UD}(T,B) = \sqrt{\frac{T\mu_0}{B}} f_{2D}\left[A\frac{T - T_c(B)}{\sqrt{TB/\mu_0}}\right],$$

where $f_{2D}$ is an unknown scaling function, and A is an appropriate constant characterizing the material. The fluctuation conductivity Δσ was determined by subtracting the normal state conductivity from the actual conductivity, calculated from resistive transitions above 65% of the normal state sheet resistance. According to this equation, the scaled experimental $\Delta\sigma(T,B)$ curves at correct values of the $T_c(B)$'s should collapse on one single curve enabling direct determination of the normal state transition $T_c$ at different magnetic fields (or $B_{c2}$ at different temperatures).

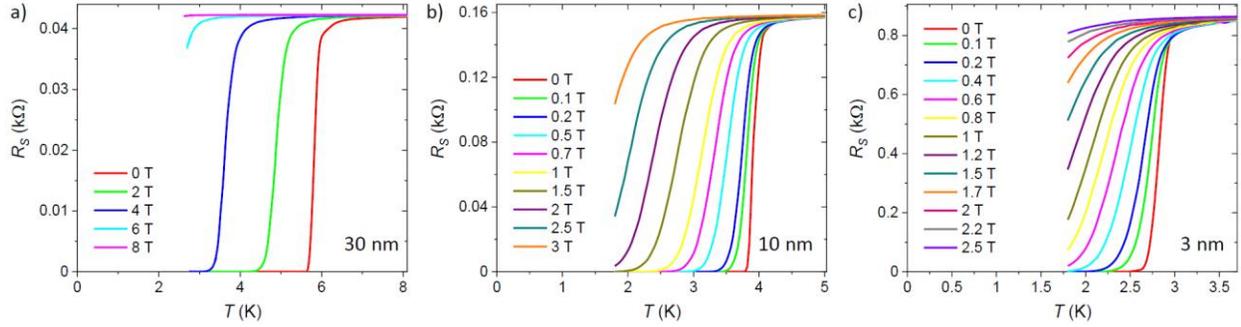

FIGURE S1. Temperature dependencies of the sheet resistance $R_S(T)$ measured in MoN 30 nm a), 10 nm b), and 3 nm c) thin films at indicated magnetic fields.

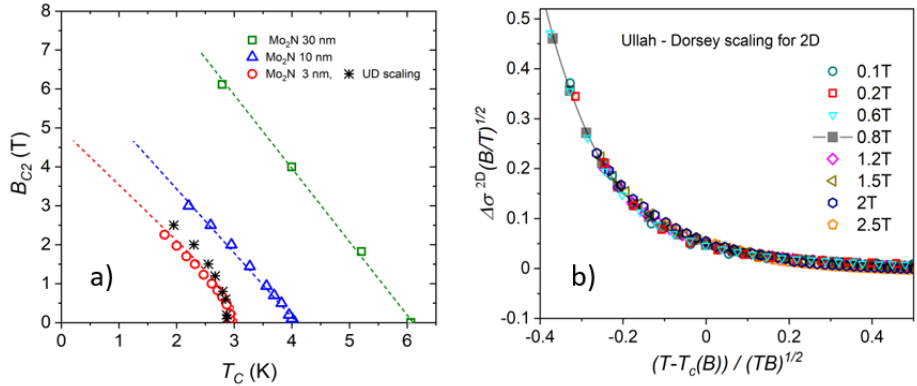

FIGURE S2. a) $B_{c2}$ of MoN films with thicknesses of 30 nm (green symbols), 10 nm (blue symbols), and 3 nm (red symbols) determined from transport data shown in Fig. S1 at 90 % of the normal state sheet resistance. The black asterisks show $B_{c2}$ values of the 3 nm thin film, determined from the Ullah-Dorsey scaling procedure for 2D superconductors. b) The Ullah-Dorsey scaling curves, calculated from the resistive transitions, shown in Fig. S1 c).

Figure S2b) shows our transport data of the 3nm thin film from Fig S1c) after succesfull UD scaling. The values of $B_{c2}(T)$ determined by this method are indicated in Figure S2a) by black asterisks. It is evident that the $B_{c2}(T)$ determined by UD scaling is close to the values determined from the $R_S(T)$ curves at the 90% transition and both criteria give a square root dependence of $B_{c2}(T)$ close to $T_c$.

**Finkel'stein's analysis of the transport data**

The critical temperatures of the thin films as the function of the corresponding sheet resistances were analyzed in the framework of the widely used Finkel'stein's model [4]. In this theory the reduced $T_c$ upon increased disorder is supposed to be the result of the increased Coulomb interaction. The decreasing $T_c$ is directly related to the increase of $R_s$ as

$$T_c = T_{c0}\exp(-1/\gamma)\{[1 + L]/[1 - L]\}^{1/\sqrt{2r}},$$

where $L = \sqrt{(r/2)}/(\gamma - r/4)$, $\gamma = 1/\ln(k_B T_{c0}\tau/\hbar)$, and $r = R_s/(2\pi^2\hbar/e^2)$, $T_{c0}$ is the $T_c$ of the bulk sample and $\tau$ is the elastic scattering time of electrons. This model enables to estimate the value of the critical sheet resistance $R_c$ at which the SIT occurs.

The level of disorder has been characterized with the Ioffe-Regel product of $k_F l$, which approaches unity near critical disorder [1]. Its value was calculated from the expression

$$k_F l = \frac{\hbar(3\pi^2)^{2/3}}{e^{5/3}}\left[\frac{R_H^{1/3}}{R_S t}\right].$$

The Hall coefficient $R_H$ was determined from Hall measurements realized in a Quantum Design Physical Properties Measurement System (PPMS).

## STM topograghy measurements

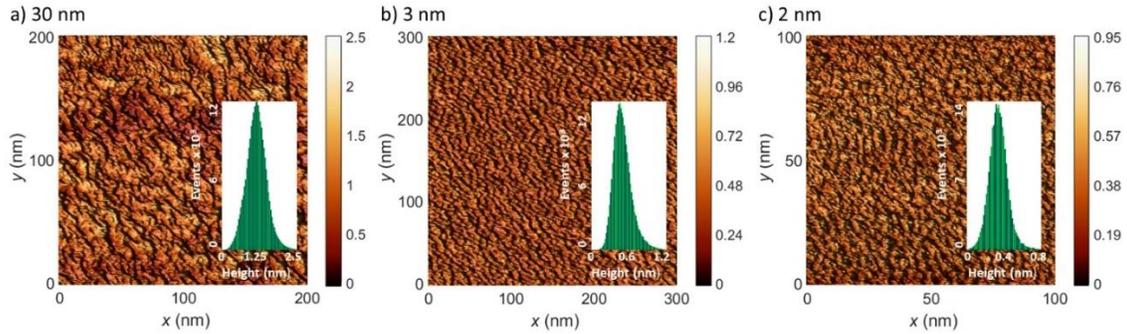

FIGURE S3. STM surface topographies of the MoN samples were measured at 30, 3 and 2 nm thicknesses in respective a), b) and c). The image sizes were 200 x 200 nm² a), 300 x 300 nm² b), and 100 x 100 nm² c). The measuring parameters are: $V = 50$ mV, $I = 1.1$ nA, $T = 1$ K a), $V = 42$ mV, $I = 3.4$ nA, $T = 0.45$K b), and $V = 90$ mV, $I = 0.5$ nA, $T = 0.5$K c). The insets show graphs of the the corrugation analysis, with 1.2 nm averaged corrugation for the 30 nm sample a) 0.5 nm for the 3 nm sample and 0.35 nm for the 2 nm sample, resp.

## Temperature dependencies of the superconducting energy gap

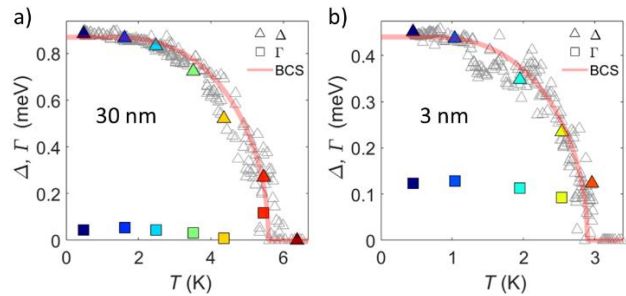

FIGURE S4. Temperature dependencies of the superconducting energy gap $\Delta(T)$ of 30 nm a) and 3 nm b) thin MoN films (open symbols) determined from fitting to Dynes formula. The color symbols are fitting values of $\Delta$ and $\Gamma$ of the curves shown in the lower panels of Fig. 2 in the main text. The red lines show predictions of the BCS theory. The fitting parameters are: $\Delta(0) = 0.87$ meV, $T_c = 5.6$ K, $2\Delta(0)/k_BT_C = 3.6$ for the 30 nm and $\Delta(0) = 0.44$ meV, $T_C = 2.9$ K, $2\Delta(0)/k_BT_C = 3.55$ for 3 nm sample.

## Magnetic field dependencies of the tunneling spectra

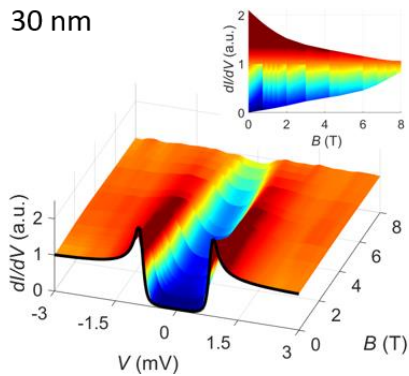

FIGURE S5. a) 3D plot of the magnetic field dependence of the typical tunneling spectra measured on 30 nm thin MoN film. The inset show the side view of the same dependence. This figure is a direct supplement to Fig. 3 in the main text.

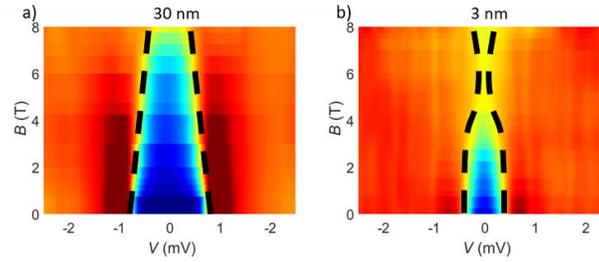

FIGURE S6. Top views of the magnetic field dependencies of the tunneling spectra presented in Fig. S5 and Fig. 3a) of the main text for 30 nm a) and 3 nm b) MoN thin films. The dashed lines emphasize the evolution of the gap-peak position with increased magnetic field.

**Theoretical magnetic field dependence of the Altshuler-Aronov effect in 3 nm thin MoN film**

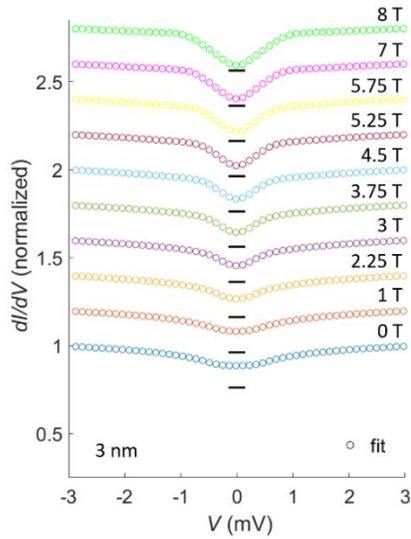

FIGURE S7. Theoretical magnetic field dependence of the Altshuler-Aronov effect in 3 nm thin MoN film used at the fitting procedure of the magnetic field dependence shown in Fig. 3a) of the main text. The short black horizontal lines below the curves show the value of 0.75.